\documentclass[a4paper,10pt,twoside]{cpc-hepnp}
\usepackage{multicol}
\usepackage{graphicx}
\usepackage{booktabs}
\usepackage{amssymb,bm,mathrsfs,bbm,amscd}
\usepackage[tbtags]{amsmath}
\usepackage{lastpage}

\begin{document}

\fancyhead[c]{\small Submitted to 'Chinese Physics C'}
\fancyfoot[C]{\small 010201-\thepage}

\title{Application of a multiscale maximum entropy image restoration algorithm to HXMT observations\thanks{Supported by the Strategic Priority Research Program on Space Science, the Chinese Academy of Sciences under grant No. XDA04010300 and National Natural Science Foundation of China (NSFC) under grant No. 11403014 (55555555) }}

\author{%
     Ju Guan$^{1}$\email{jguan@ihep.ac.cn}%
\quad Li-Ming Song$^{1}$
\quad Zhuo-Xi Huo$^{2,3}$
}
\maketitle

\address{%
$^1$ Institute of High Energy Physics, Chinese Academy of Sciences, Beijing 100049, China\\
$^2$ Qian Xuesen Laboratory of Space Technology, China Academy of Space Technology, Beijing 100094, China\\
$^3$ Department of Physics, Tsinghua University, Beijing 100084, China\\
}

\begin{abstract}
This paper introduces a multiscale maximum entropy (MSME) algorithm for image restoration of the Hard X-ray Modulation Telescope (HXMT), which is a collimated scan X-ray satellite mainly devoted to a sensitive all-sky survey and pointed observations in the 1-250 keV range. The novelty of the MSME method is to use wavelet decomposition and multiresolution support to control noise amplification at different scales. Our work is focused on the application and modification of this method to restore diffuse sources detected by HXMT scanning observations. An improved method, the ensemble multiscale maximum entropy (EMSME) algorithm, is proposed to alleviate the problem of mode mixing exiting in MSME. Simulations have been performed on the detection of the diffuse source Cen A by HXMT in all-sky survey mode. The results show that the MSME method is adapted to the deconvolution task of HXMT for diffuse source detection and the improved method could suppress noise and improve the correlation and signal-to-noise ratio, thus proving itself a better algorithm for image restoration. Through one all-sky survey, HXMT could reach a capacity of detecting a diffuse source with maximum differential flux of 0.5 mCrab.
\end{abstract}



\footnotetext[0]{\hspace*{-3mm}\raisebox{0.3ex}{$\scriptstyle\copyright$}2013
Chinese Physical Society and the Institute of High Energy Physics
of the Chinese Academy of Sciences and the Institute
of Modern Physics of the Chinese Academy of Sciences and IOP Publishing Ltd}%

\begin{multicols}{2}

\section{Introduction}

In the hard X-ray band, photons with energies from tens of keV to a few hundred keV can neither be focused nor can their arrival direction be determined from Compton scattering or e$^{\pm}$ pair production~\cite{lab1}. Thus imaging in the hard X-ray band is mainly achieved by various modulation techniques, such as the coded-mask technique and collimator, leading to coded-mask aperture telescopes and collimated telescopes. Because the observed data of these modulation instruments records just the correlation between the sky image and an array derived from the instrument properties, one must reconstruct the sky image from the observed data with numerical methods. There exist many mature analysis methods and a lot have been investigated in-depth in hard X-ray astronomy. They can be divided into four catalogs: linear methods, statistical methods, iterative methods, and wavelet-based methods~\cite{lab2}. Cross correlation~\cite{lab3} proves to be the simplest and most convenient linear method for detecting point sources. But as it makes no deconvolution of the observed data, it might not be suitable for diffuse source restoration. The statistical methods (such as the maximum entropy method~\cite{lab4} and maximum likelihood method~\cite{lab5}) provide a statistical solution to the deconvolution problem. But as all the information of the source and observation has been degenerated into one statistic, they output restored images with limited resolution and sensitivity~\cite{lab6}, especially for point source detection. To overcome this problem, the Direct Demodulation method~\cite{lab6,lab7} has been proposed, which is a good representation of the iterative methods. By solving the observation equation and implementing physical constraints on each iterative output, this method offers much higher resolution images and is flexible to restore images with both point sources and diffuse sources. Wavelet-based methods are a kind of method combining wavelet and traditional methods. For instance, the multiscale maximum entropy (MSME) method, proposed in 1996~\cite{lab8}, is based on the maximum entropy method. It uses wavelets to decompose an image into different frequency bands and applies multiresolution support and different smoothness constraints to each band to control noise amplification, leading to a better restoration for high and low spatial frequency structures at the same time.

The Hard X-ray Modulation Telescope (HXMT) is a collimated scan X-ray satellite~\cite{lab1} with three slat-collimated detectors on board: the High Energy X-ray telescope (HE), the Medium Energy X-ray telescope (ME) and the Low Energy X-ray telescope (LE). It will perform scanning observations (i.e. an all-sky hard X-ray survey, deep imaging observations of selected sky regions) and pointed observation in the 1-250 keV range~\cite{lab9}. Many simulations have confirmed its high sensitivity and angular resolution for point source detection in scanning observation mode~\cite{lab1,lab6,lab9,lab10,lab11}, but few studies have been done for diffuse sources. However, diffuse source detection is of great importance in astronomy. Many sources of interest must be considered as diffuse in order to properly derive their morphologies and intensity, such as remnants, diffuse interstellar emission from various high-energy processes and the cosmic X-ray background. Comparing the key features of the telescope (i.e. large effective area and field of view of degree order) with other contemporary satellites, we find that HXMT does have an advantage in detecting diffuse sources and large scale structures spanning several to tens of degrees, which are difficult to obtain by imaging telescopes because of their small rigidity factor and ineffectiveness above high energy, and difficult to obtain by coded aperture telescopes because of their complicated image distortions~\cite{lab1}. (Indeed, studies have revealed the sensitivity of any coded aperture telescope degrades almost linearly with the source extent~\cite{lab12}.) Thus, HXMT is hoped to fulfill the gap and bring forth abundant knowledge of diffuse emission in the hard X-ray band. But before realizing the goal, we need powerful tools to extract diffuse sources from data. As the MSME method is very suitable for simultaneously restoring high and low spatial frequency structures in an image, we will investigate it in this paper.

In Section 2, we will give a brief introduction of this method, and some modifications which have been made to fit for HXMT data. An improved method based on the MSME method has been proposed to give better restoration. Simulations and results are presented and compared in Section 3. The main discussions and conclusions about this work are given in the last section.

\section{Image restoration methods}

The relation between the observational data $I(x,y)$ and the intensity distribution $O(x,y)$ of a sky region can be described by the 
following observation equation:
\begin{eqnarray}
\label{eq1}
I(x,y)=(O*P)(x,y)+N(x,y).
\end{eqnarray}
where $P(x,y)$ is the Point Spread Function (PSF) of the telescope, and $N(x,y)$ is additive noise introduced during the observation. The image restoration problem is to determine $O(x,y)$ from known $I(x,y)$ and $P(x,y)$. 

\subsection{Multiscale maximum entropy method}

\subsubsection{Definition of multiscale maximum entropy method}

The maximum entropy method (MEM) provides a statistical solution to the problem by minimizing the target function~\cite{lab13}

\begin{eqnarray}
\label{eq2}
J(O)=\sum_{pixels}\frac{(I-P*O)^2}{2{\sigma_I}^2}-{\alpha}S(O)=\frac{\chi^2}{2}-{\alpha}S(O).
\end{eqnarray}
where $\alpha$ is a parameter that fixes the relative weight between chi-squared and the entropy $S(O)$. But as the entropy is global quantity calculated on the whole image $O$, it is difficult to find a perfect value of $\alpha$ when the observed image has simultaneously high and low spatial frequency structures. Therefore, Pantin and Starck~\cite{lab8} have proposed the multiscale maximum entropy (MSME) method. The novelty of the method is to use wavelet transforms to decompose an image into different frequency bands and use a multiresolution support to separate signal and noise, thus solving the problem of MEM to choose $\alpha$. The mathematical realization is below by defining the entropy as

\begin{eqnarray}
\label{eq3}
S(O)=\frac{1}{\sigma_I}\sum_{scalesj}\sum_{pixels}A(j,x,y)\sigma_j(w_j(x,y)-m_j\nonumber\\-{\vert}w_j(x,y){\vert}\ln\frac{{\vert}w_j(x,y){\vert}}{m_j}).
\end{eqnarray}
where $\sigma_I$ is the standard deviation of the noise in the image $I$, and $\sigma_j$ the noise standard deviation at scale $j$. $w_j$ are wavelet coefficients, while $m_j$ represents the value of wavelet coefficients in the absence of any input signal~\cite{lab8}. The A function of the scale $j$ and the pixels $(x,y)$ is $A(j,x,y)=1-M(j,x,y)$, with the multiresolution support $M(j,x,y)$ defined as:

\begin{equation}
\label{eq4}
M(j,x,y)=
\begin{cases}
1 &\text{, if $w_j(x,y) \geqslant k\sigma_j$} \\
0 &\text{, if $w_j(x,y) < k\sigma_j$}
\end{cases}
\end{equation}
It is introduced to distinguish signal and noise. The parameter $k$ fixes the threshold level and $k=3$ is generally adopted in astrophysics~\cite{lab8}. Thus $A(j,x,y)$ can be seen as a parameter determining the degree of regularization: strong regularization is implemented when $A$ is near 1 while regularization is weak when $A$ is around 0. 

The minimization of Eq.~(\ref{eq2}) can be implemented using the steepest descent method, which gives the iterative scheme:

\begin{equation}
\label{eq5}
O^{n+1}=O^n-{\gamma}{\bigtriangledown}(J(O^n))
\end{equation}
where $\gamma$ is the step size while ${\bigtriangledown}(J(O^n))$ is the gradient of $J(O^n)$.

\subsubsection{Modification of multiscale maximum entropy method}

Since the data in our simulation case is a little different from that required by MSME, two preparations have to be done when introducing the MSME method for HXMT data analysis.

The first is the noise type of the data. The noise in our simulation is Poisson type, while the default noise in MSME is Gaussian. So we have to transform our data to the Gaussian case before restoration. The second is that we have to restore the sky image from 15 detected images, while the MSME is originally designed to restore it from one detected image. Therefore, modification should be made to use all the observed data to figure out the sky image. We solve this problem by modifying Eq.~(\ref{eq5}) as

\begin{equation}
\label{eq6}
O^{n+1}=O^n-\frac{1}{N}\sum_{i=1}^N{\gamma_i}{\bigtriangledown}(J_i(O^n))
\end{equation}
i.e. using the average gradient of the 15 detectors as the final gradient. N is the number of detectors.

The steps for MSME are as follows:

1) transform the observed data to the Gaussian case using Anscombe transform~\cite{lab8}.

2) use modified MSME method to restore the data, with each iterative result applying background constraint, and the iteration stopping when it meets the stopping criterion.

3) transform the result to Poisson case using inverse-Anscombe formula.

Applying the background constraint to each iterative result is a technique learned from the Direct Demodulation method, in which the background constraint plays an important role in suppressing noise and controlling the iterative process to produce a satisfactory reconstruction, especially for cases with poor statistics. The background constraint is enforced as:

\begin{equation}
\label{eq7}
O(x,y)=
\begin{cases}
O(x,y)&\text{, if $O(x,y) \geqslant B$} \\
B&\text{, if $O(x,y) < B$}
\end{cases}
\end{equation}
where $B$ is the estimated background and can be derived from the observed data (which has been transformed to the Gaussian case) by an iterative algorithm similar to that done for estimating the $\sigma_I$\cite{lab14}. The algorithm uses  multiresolution support to find a set of pixels $Q$ in the image which are due only to the noise, and calculates the standard deviation from them as $\sigma_I$. As a pixel belonging to the set $Q$ is defined as pixel with $M(j,x,y)=0$ for all $j$ (i.e. no significant coefficient at all scales), we can equivalently say that it belongs to the background. More specifically, the high frequency parts of these pixels relate to noise, while the low frequency part represents  information on the background. Therefore the background B could be estimated as the mean of the low frequency parts of these pixels. The algorithm is below:

1) estimate the standard deviation of the noise in $I$ (which has been transformed to the Gaussian case) to get $\sigma_I^{(0)}$

2) decompose $I$ with $\grave{a}$ trous algorithm:

\begin{equation}
\label{eq8}
I(x,y)=c_p(x,y)+\sum_{j=1}^pw_j(x,y)
\end{equation}
where $w_j$ are wavelet coefficients corresponding to the high frequency parts of $I$, while  $c_p$ is the low frequency part where noise is negligible.

3) $n=0$

4) compute the multiresolution support M from $w_j$ and $\sigma_I^{(n)}$

5) select all the pixels that satisfy $M(j,x,y)=0$ for all $j$ to form the set $Q$

6) calculate the standard deviation $\sigma_I^{(n+1)}$ of $I(x,y)-c_p(x,y)$ for all the pixels in $Q$. 

7) $n=n+1$

8) repeat step (4-7) until $\frac{{\vert}\sigma_I^{(n)}-\sigma_I^{(n-1)}{\vert}}{\sigma_I^{(n)}}<\varepsilon$

9) compute the mean of $c_p(x,y)\in Q$ as the background $B$

\subsection{Ensemble multiscale maximum entropy method}

An important aspect of image restoration processing is that of the treatment of additive noise introduced by the observation. Usually most data analysis methods are designed to remove noise, however, there are cases when noise is added to help suppress noise and obtain better restoration results.

MSME belongs to the former class. The novelty of it is to use wavelet decomposition and multiresolution support to control noise amplification. But as the $\grave{a}$ trous wavelet may not be the exact base for all frequency structures, the MSME may encounter the mode-mixing problem, i.e. noise and signal cannot be separated completely from each other at any single scale. Thus, if there exists a residual component of noise in the signal, it would not be regularized due to its being above the regularization threshold. Moreover, the precision of the estimated $\sigma_I$ and then that of the regularization threshold and the estimated background would be affected. To overcome this problem, we seek inspiration from the latter kind of methods. We introduce a noise assisted data analysis technique to the MSME method just like that done for the Empirical Mode Decomposition (EMD) method~\cite{lab15}, which is an adaptive time-frequency data analysis method also facing the mode-mixing problem. We name the improved method the  ensemble multiscale maximum entropy (EMSME) because it consists of restoring an ensemble of white noise-added images by MSME and treats the ensemble mean as the final true result.

The steps for EMSME method are as follows:

1) estimate the noise level of the original observed data.

Since the background noise in our observed data is due to Poisson fluctuation of the background, the estimation of the noise level is equivalent to estimating the background of the original observed data. The background estimation is similar to that in the MSME but with the mean calculated on the original observed data (i.e. Poisson type).

2) generate white noise with the same distribution as the estimated noise.

In our simulaton we do it just by Poisson sampling from the estimated background.

3) add the white noise to the original observed data to form new data.

4) restore the new data using the modified MSME method.

5) repeat step (2-4) many times, and treat the mean of all restored images as the final result.

We expect the EMSME method to use the full advantage of the statistical characteristics of white noise. It perturbs the observed data and alleviates the coupling between signal and original noise, thus solving the mode-mixing to some extent, and cancels itself out in the space ensemble mean after serving its purpose.

\subsection{Evaluation criteria for image restoration quality}

The quality of a restored image can be evaluated through visual inspection and quantitative criteria~\cite{lab16}. 

A classical criterion is given by the correlation coefficient between the original image $O(x,y)$ and the restored one $\tilde{O}(x,y)$, which is defined as: 

\begin{equation}
\label{eq10}
C_{or}=\frac{\sum_{x=1}^N\sum_{y=1}^NO(x,y)\tilde{O}(x,y)}{\sum_{x=1}^N\sum_{y=1}^NO^2(x,y)\sum_{x=1}^N\sum_{y=1}^N{\tilde{O}}^2(x,y)}
\end{equation}

Another way to compare two images is to determine the Signal-to Noise Ratio (SNR).

\begin{equation}
\label{eq11}
SNR_{dB}=10\lg\frac{\sum_{x=1}^N\sum_{y=1}^NO^2(x,y)}{\sum_{x=1}^N\sum_{y=1}^N(O(x,y)-\tilde{O}(x,y))^2}
\end{equation}

\section{Simulation and results}

\subsection{Configuration of simulation}

\subsubsection{Building the input model}

We choose Centaurus A (Cen A for short) as our simulated object. As the nearest radio galaxy to us, its proximity (D=3.8$\pm$0.1 Mpc~\cite{lab17}) makes it an attractive target for comprehensive physical studies. The most prominent features of Cen A are its two large radio lobes, which extend for $\sim$600 kpc ($\sim$10 degrees) in the north-south direction and are  $\sim$250 kpc ($\sim$4 degree) wide~\cite{lab18}. Their $\gamma$-ray glow emission has been detected by the Fermi Gamma-ray Space Telescope~\cite{lab19}, providing insights into radiating high-energy particles in the lobes. However, the results on X-ray detection are mixed. Handcastle~\cite{lab20} argued that X-ray emission fills the field of view of modern soft X-ray imaging instruments, making its detection difficult (if at all possible). On the other hand, the wide field of view instruments like INTEGRAL and Swift have limited sensitivity to extended emission. With improved sensitivity, HXMT also has a large field-of-view and will thus offer an opportunity to study highly extended X-ray sources. We believe that Cen A would be an important target for HXMT, so have chosen it for this work.

Since the lobes of Cen A have not been detected in the X-ray band, we extrapolated the measured Fermi/LAT gamma-ray spectrum~\cite{lab19}, assuming it follows a power law, to the 50-250 keV band. A wavelet-Lucy method has been introduced to make the morphology smoother. Here, we focus on the lobes of Cen A, so we excluded an inner core region of radius $1^{\circ}$. A fast adjacent inpainting algorithm was used to interpolate the value of the core region from the outer region. Fig.~\ref{fig1} shows the final input model. The maximum differential flux is $2.5\times10^{-4}$ cts/s/cm$^2$, corresponding to 2.5 mCrab. (A 1-mCrab source causes a total count rate of $10^{-4}$ cts/s/cm$^2$ for HE at energy band of 50 keV-250 keV.)

\subsubsection{Setting the observation conditions}

We investigate the performance of the High Energy X-ray telescope (HE), which is the main payload onboard HXMT. Moreover, we take into account only the 15 detectors with narrow field of view ($1^{\circ}\times5^{\circ}$ at FWHM). They are divided into three groups with orientation of a cross angle of $60^{\circ}$ to one another (Fig.~\ref{fig2}). Each detector has an effective  area of 277.8 cm$^2$, and the in-orbit background counts is 39 cts/s, which is derived from the latest background study of HXMT~\cite{lab21}. The energy band is 50 keV-250 keV. The scanning step is $0.2^{\circ}$, which is feasible in the observation mode of HXMT, and the observation time of each scanning point is 3 seconds, which is determined from the all-sky survey mode.

\begin{center}
	\includegraphics[width=7cm]{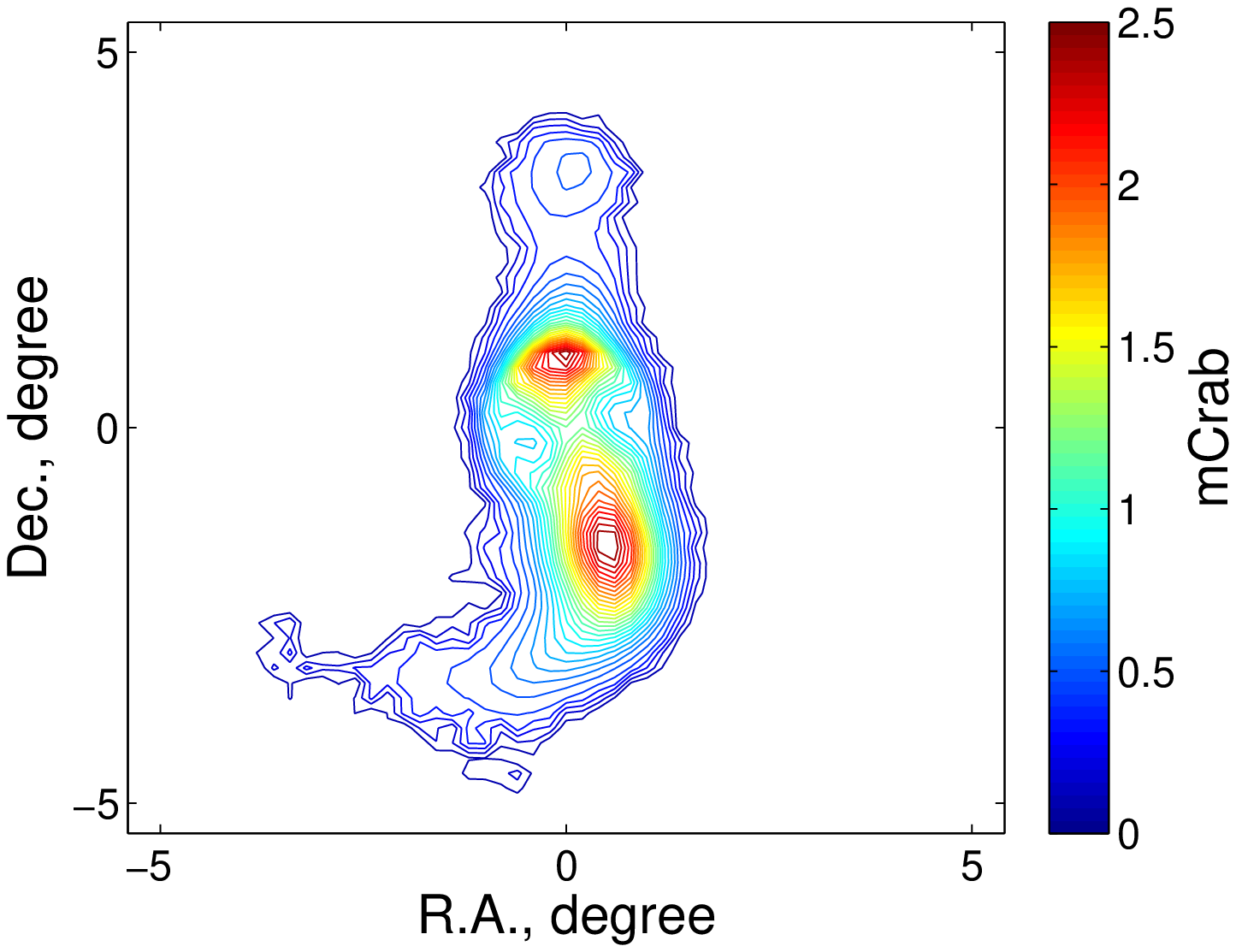}
	\figcaption{\label{fig1}   Input image centered at the core of Cen A. The coordinate R.A. is short for right ascension, while Dec. stands for declination. }
\end{center}

\begin{center}
	\includegraphics[width=7cm]{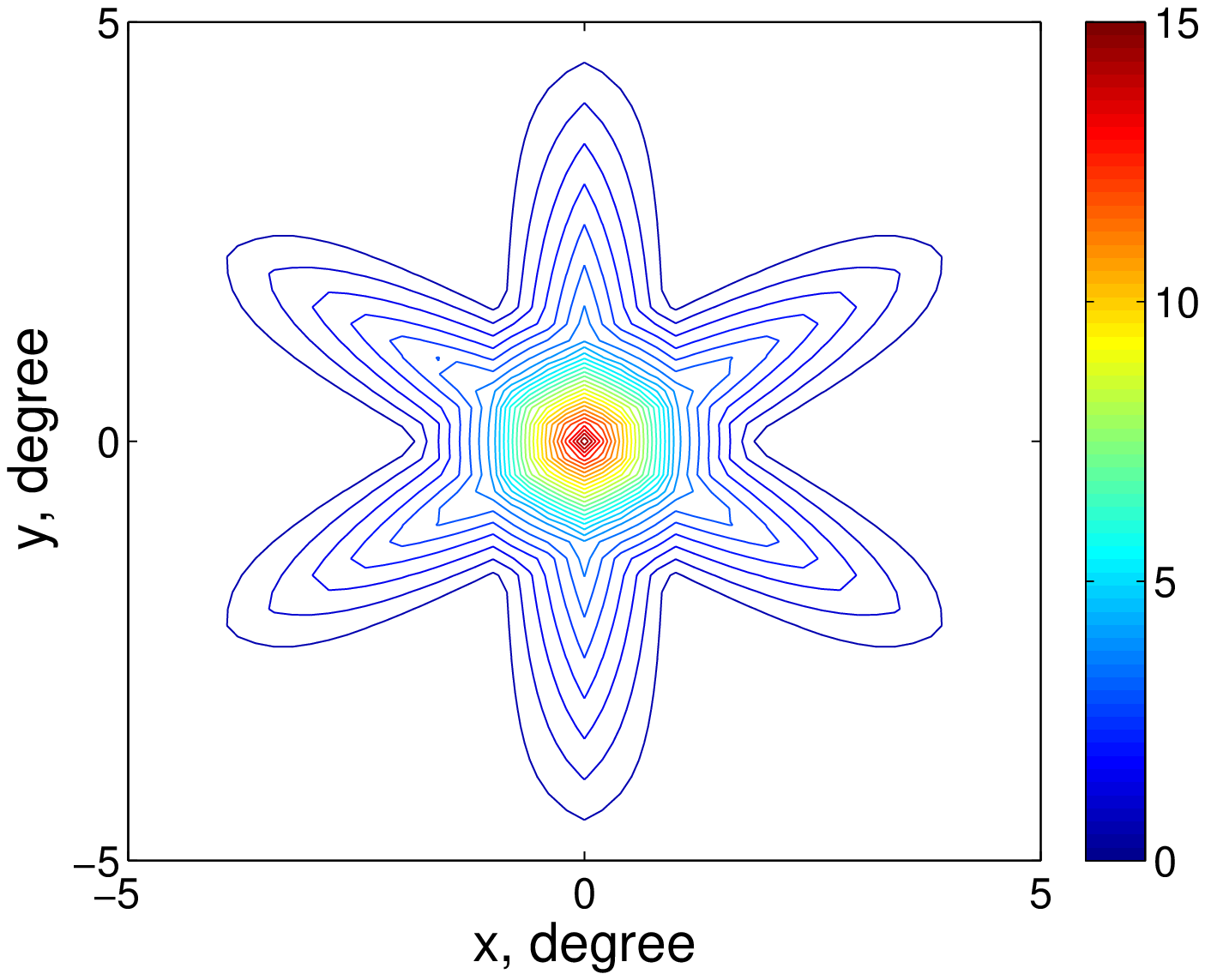}
	\figcaption{\label{fig2}   Overlay of PSF of the 15 detectors, with x and y the Cartesian coordinates set up on the detector plane.}
\end{center}

\subsubsection{Obtaining the observed data}

The observed data of the scanning survey in a $\sim10^{\circ}\times10^{\circ}$ region with Cen A in the center is generated by Monte-Carlo simulations. The scan mode is: the angular difference between two consecutive pointings is $0.2^{\circ}$ and each pointing lasts 3 s (corresponding to exposure time of 375 s for Cen A). The counts collected by the detectors at each pointing consist of two components: photons from Cen A and that from the instrument background, both with the Poisson fluctuation taken into consideration. Fig.~\ref{fig3} shows the observed data of the 15 detectors individually.

\begin{center}
	\includegraphics[width=8cm]{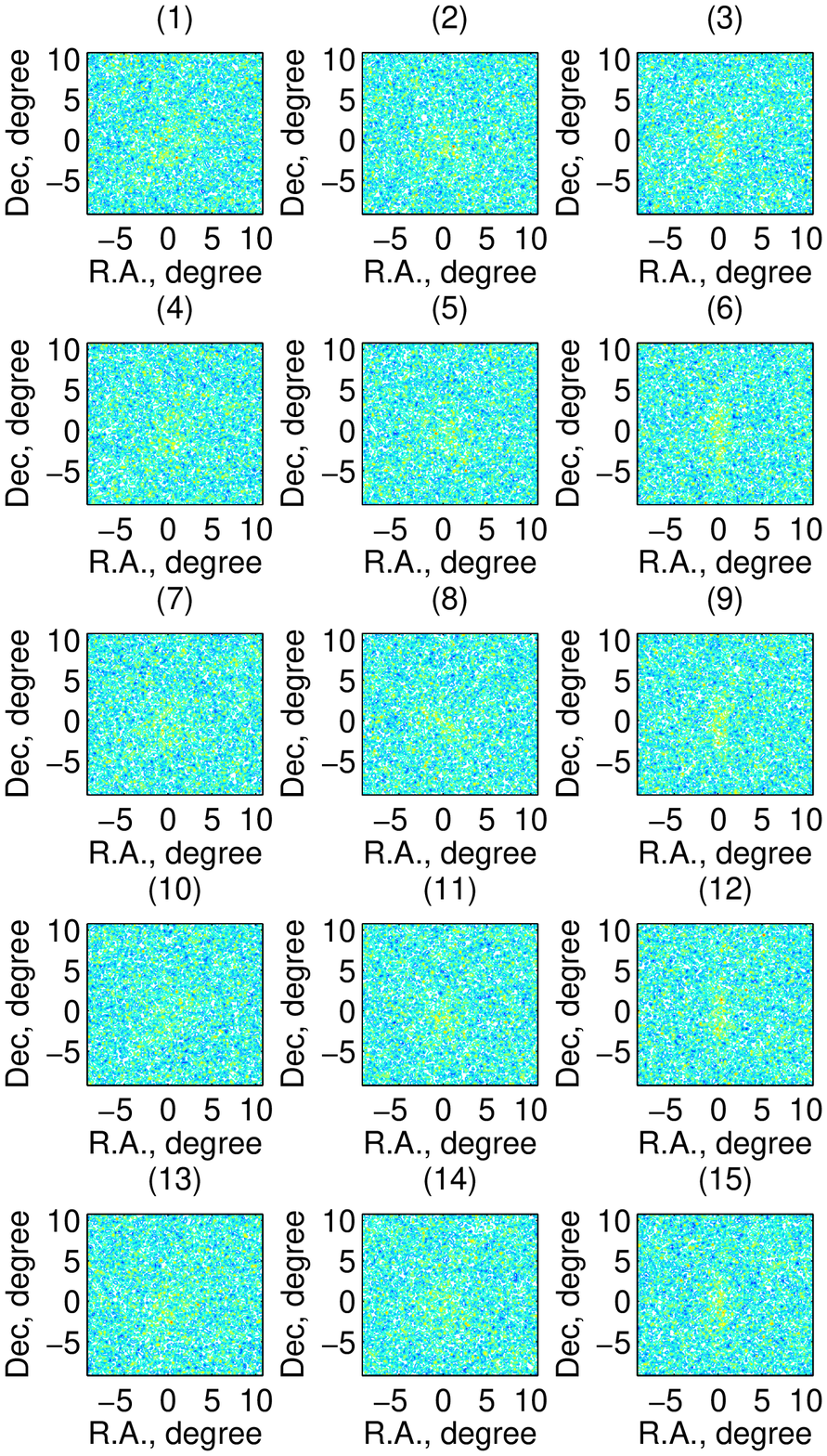}
	\figcaption{\label{fig3}   The observed data of the 15 detectors. The color bar for all images is from 80 to 180 cts. }
\end{center}

\subsection{Simulation results}

We have implemented these two methods for the restoration of (the simulated data of) Cen A. The results are evaluated through visual impression and some quantitative criteria.

\subsubsection{The iterative stopping criterion}

As the restoration algorithm is implemented through the iterative  process, we first have to decide the optimal stopping condition that will output the best restored result. We investigate three parameters as iterative stopping criteria, i.e. the standard deviation, the skewness, and the total variance of the residual between two successive restored images. Fig.~\ref{fig4} shows their changes with the iterative step. From these curves, we can see that the skewness reveals a conspicuous feature which goes down then rises and finally oscillates. We investigate 4 points of the curve (i.e. the minimum point, zero crossing point, the maximum point before oscillation, oscillation point) through their corresponding restored images (Fig.~\ref{fig5}, Fig.~\ref{fig6}). Visual impression indicates that with increasing iterative step, the restored image becomes smoother due to noise regularization and over-reconstruction due to over-iteration. Therefore, there must be a right point in the middle that will output the best restored image (with enough noise suppression and without over-reconstruction). Indeed, this conjecture is proved by the curves of correlation coefficient and SNR, which reach maximum in the middle (Fig.~\ref{fig7}). More accurately, the best iterative stopping step is when the skewness first reach the mean of the subsequent oscillatory values.

\begin{center}
	\includegraphics[width=8cm]{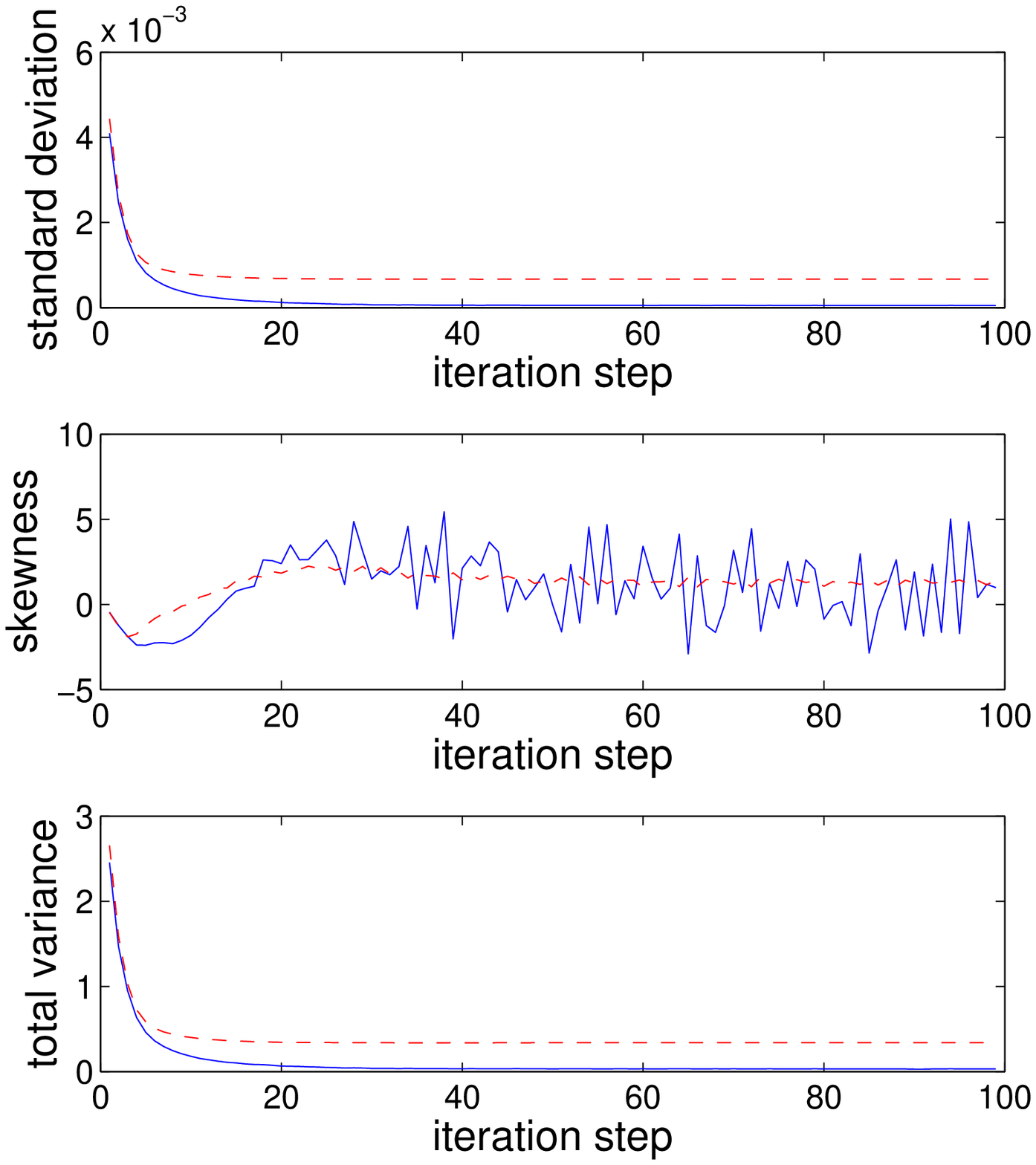}
	\figcaption{\label{fig4}   The iterative stopping criteria for MSME method (blue solid line) and EMSME method (red dashed line). It shows the standard deviation, the skewness, and the total variance of the residual between two successive restored images, respectively, varying with the iteration steps. }
\end{center}

\begin{center}
	\includegraphics[width=8cm]{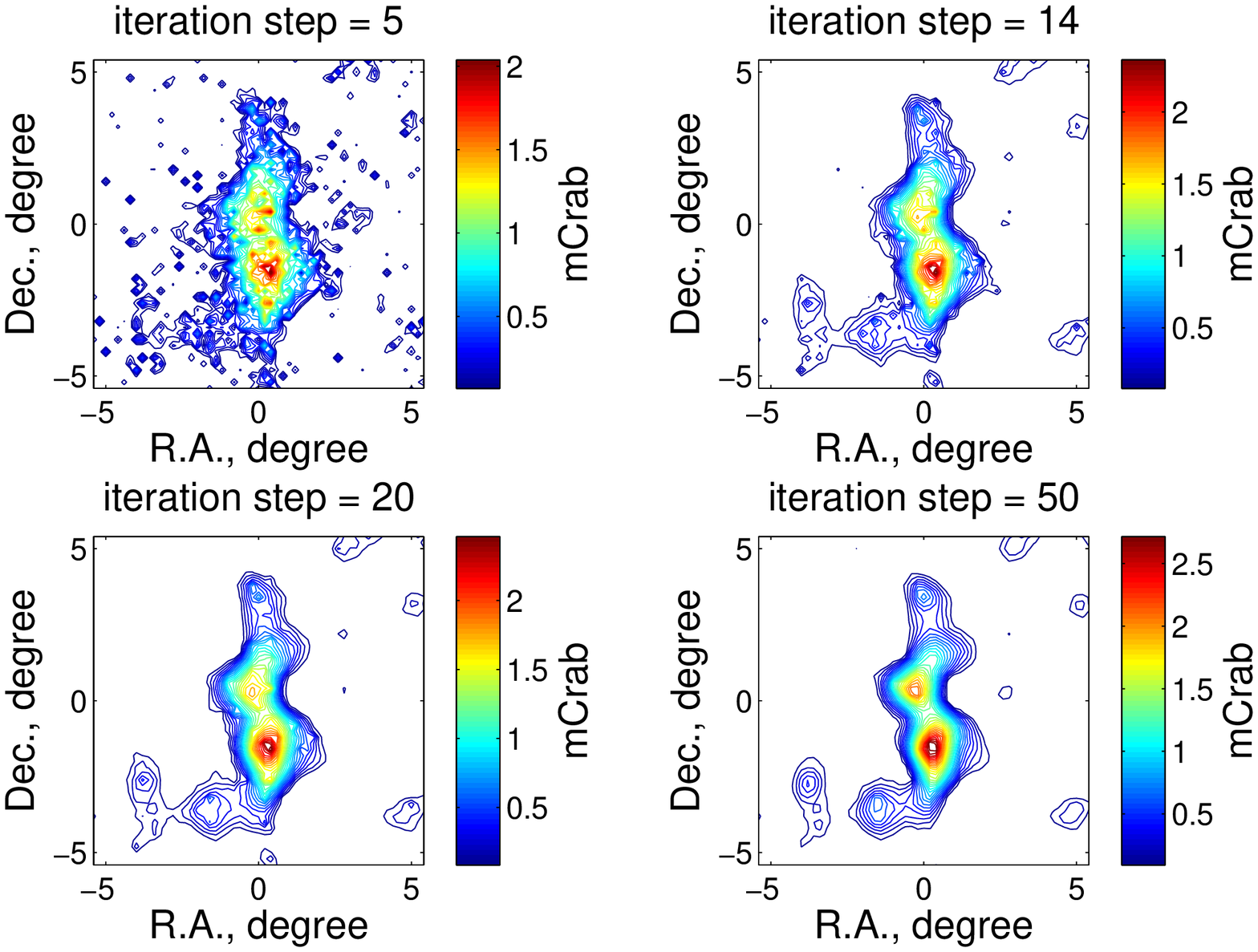}
	\figcaption{\label{fig5}   Restored images by MSME method. Their corresponding iteration steps are 5, 14, 20, 50, respectively. }
\end{center}

\begin{center}
	\includegraphics[width=8cm]{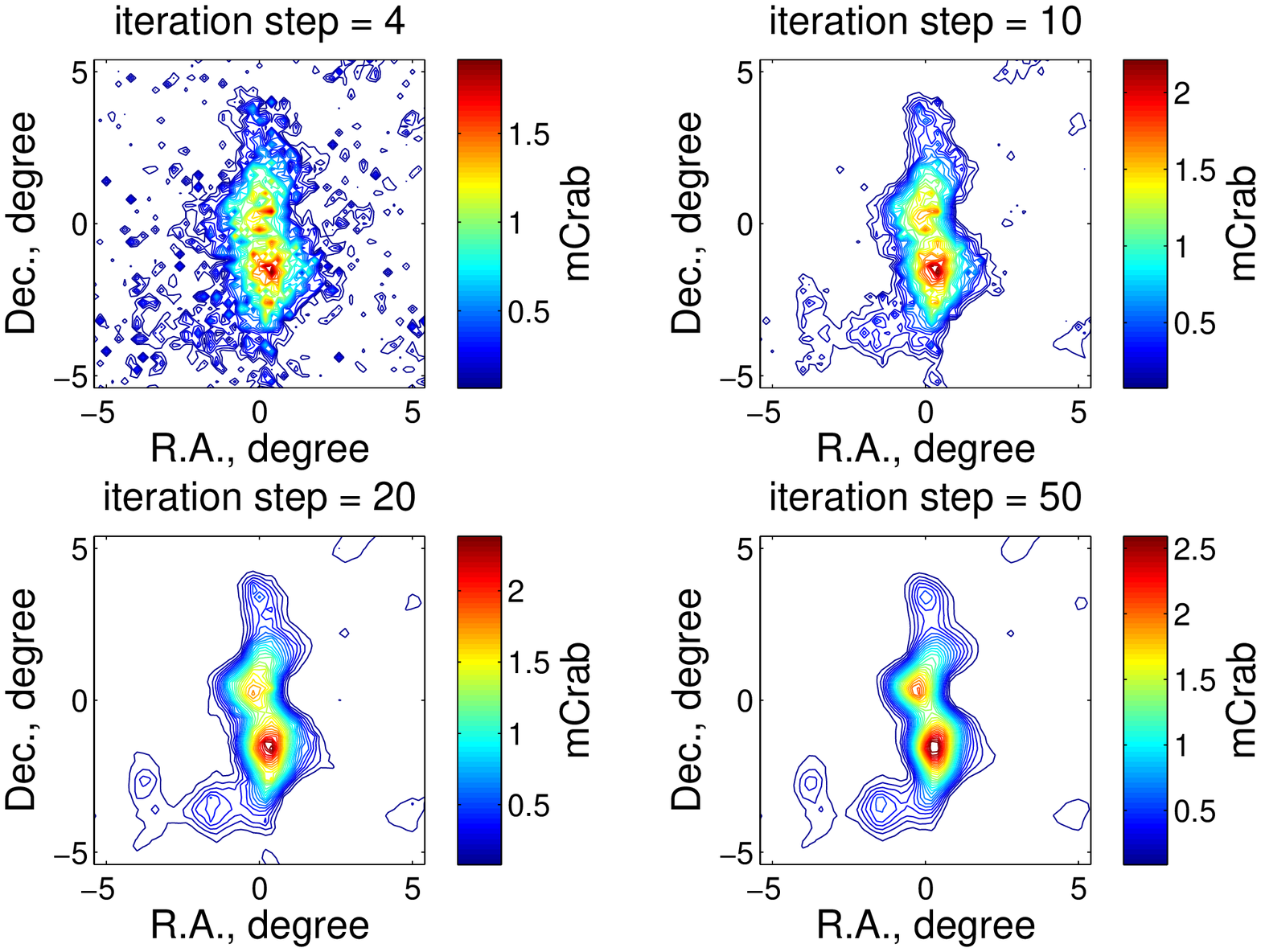}
	\figcaption{\label{fig6}   Restored images by EMSME method. Their corresponding iteration steps are 4, 10, 20, 50, respectively. }
\end{center}

\begin{center}
	\includegraphics[width=8cm]{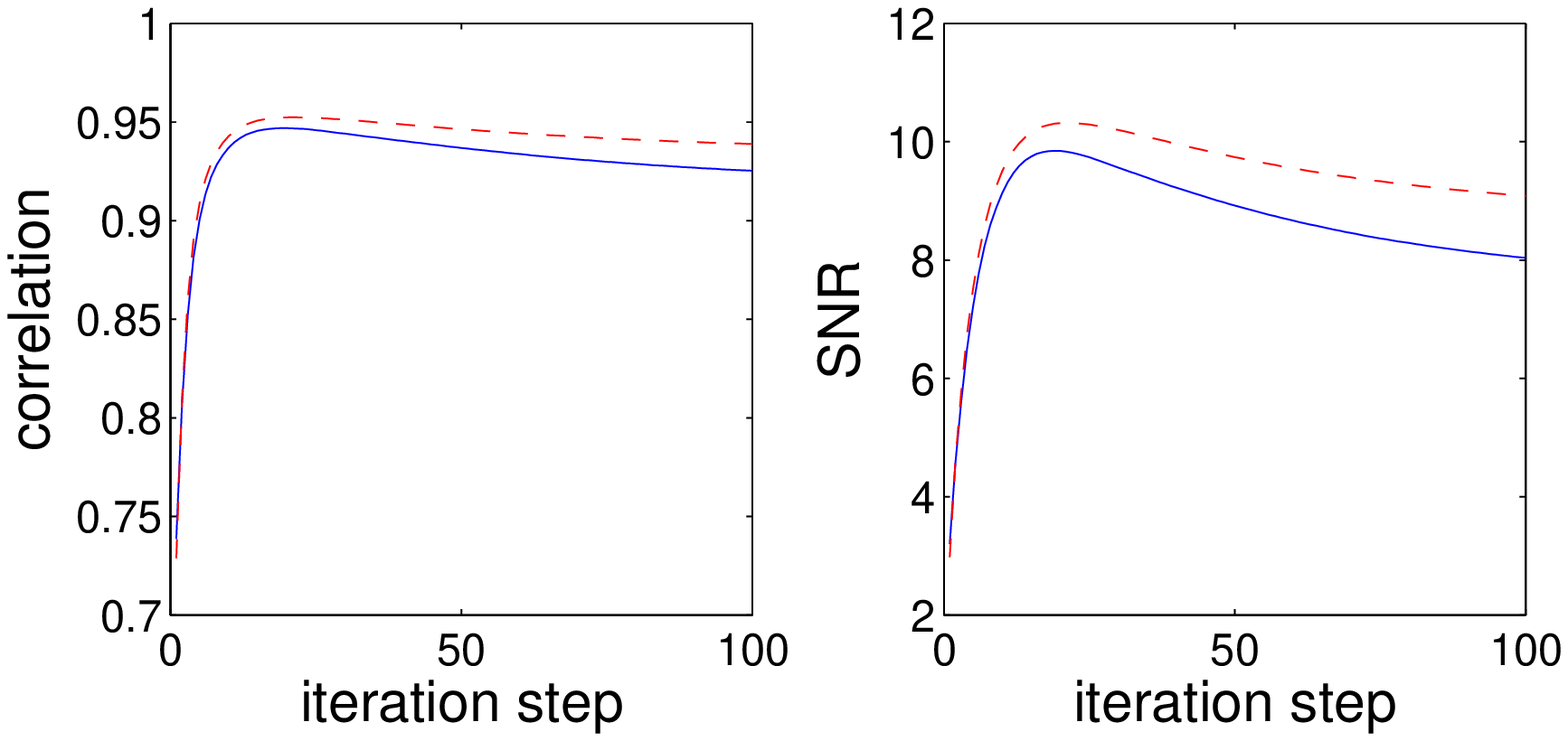}
	\figcaption{\label{fig7}   The curves of correlation coefficient (left) and SNR (right) for MSME method (blue solid line) and EMSME method (red dashed line). }
\end{center}

\subsubsection{Comparison of the two methods}

We compare the restored images of these two methods obtained at the best stopping step through morphology, distribution of the background noise (Fig.~\ref{fig8}) and some quantitative variables (Table 1). Simulation results show that the MSME method with background constraint has a powerful denoising ability, with the restored image reaching a very high quality (the correlation coefficient is 0.9468 and the SNR is 9.840), thus suitable for the restoration of diffuse X-ray sources. The EMSME method could further suppress noise according to visual impression or the distribution of the background noise, and improve the correlation coefficient (from 0.9486 to 0.9521) and SNR (from 9.840 to 10.284), thus better for image restoration. Indeed, the weakening of the background noise also means the improvement of the detection sensitivity.

To further test the properties of EMSME, we magnify the amplitude of the added white noise 1.5 times and change the number of ensemble members from 200 to 500, respectively. Simulation results indicate that the quality of restoration has little dependence on the investigated variables. Thus we suggest the amplitude of the added white noise to be the same as that in the original observed data. The number of ensemble members only needs to be large enough to cancel out the added white noise in the final ensemble mean.  

With the aid of these methods, HXMT could detect the lobes of Cen A  with substantially correct morphology and flux. The integral flux of the restored Cen A is 376 mCrab by the MSME method and 369 mCab by the EMSME method versus the input 363 mCrab.

\subsubsection{Limited detectivity of Cen A-like sources by HXMT}

To test the limited detectivity of Cen A-like sources by HXMT, we decrease the flux of the input model by 5 and 10 times. The same simulation is implemented and the comparison between the input model and the restored images are presented in Fig.~\ref{fig9} and Fig.~\ref{fig10}, respectively. Fig.~\ref{fig9} shows that the flux in the southern lobe has been reconstructed more or less correctly, while the northern lobe is obviously influenced by the observed noise because it has a larger flux than the southern lobe. The situation is worse for fainter source restoration. A discrete source could be split into several sources (Fig.~\ref{fig10}) due to the worse statistics of the observed data. At a result, based on the maximum differential flux of the southern lobe in Fig.~\ref{fig9}, we roughly give the limited detectivity of Cen A-like sources by HXMT to be 0.5 mCrab.

\end{multicols}

\ruleup

\begin{center}
	\includegraphics[width=16cm]{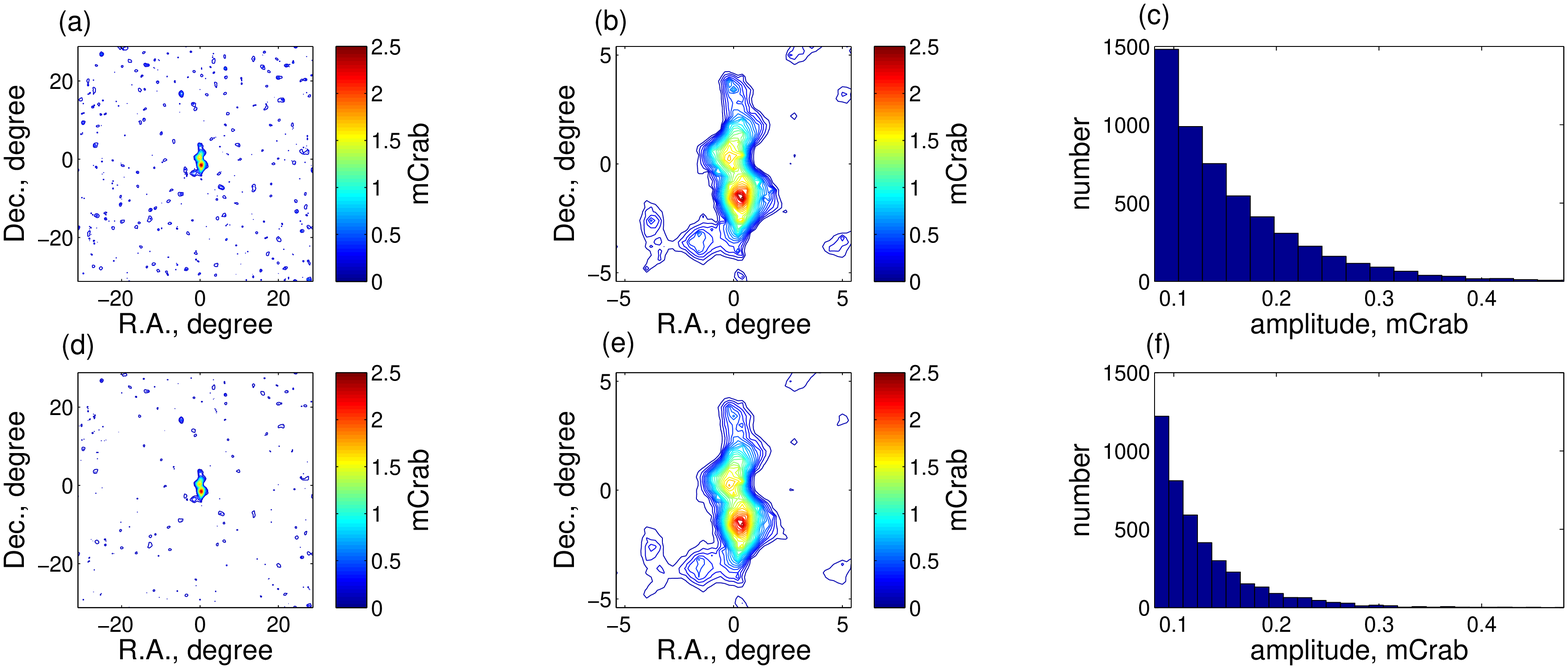}
	\figcaption{\label{fig8}   Comparison of the two methods. (a) restored image by MSME method; (b) a larger view of (a); (c) the distribution of the background noise of (a); (d) restored image by EMSME method with 200 ensemble members; (e) a larger view of (d); (c) the distribution of the background noise of (d)}
\end{center}

\begin{center}
	\tabcaption{ \label{tab2}  Comparison between the methods used.}
	\footnotesize
	\begin{tabular*}{170mm}{@{\extracolsep{\fill}}ccccccc}
		\toprule Methods & Integral flux/mCrab   & Correlation coefficient  & SNR  & Background distribution/mCrab \\
		\hline
		Restored image by MSME method  &  376  &  0.9468             & 9.840          & 0.155$\pm$0.073\\
		Restored image by EMSME method  & 369  &  0.9521             & 10.284         & 0.127$\pm$0.050\\
		Input image                     & 363  &  1\hphantom{00000}  & $\infty$       &  \\
		\bottomrule
	\end{tabular*}%
\end{center}

\ruledown

\vspace{5mm}

\begin{multicols}{2}

	\begin{center}
		\includegraphics[width=9cm]{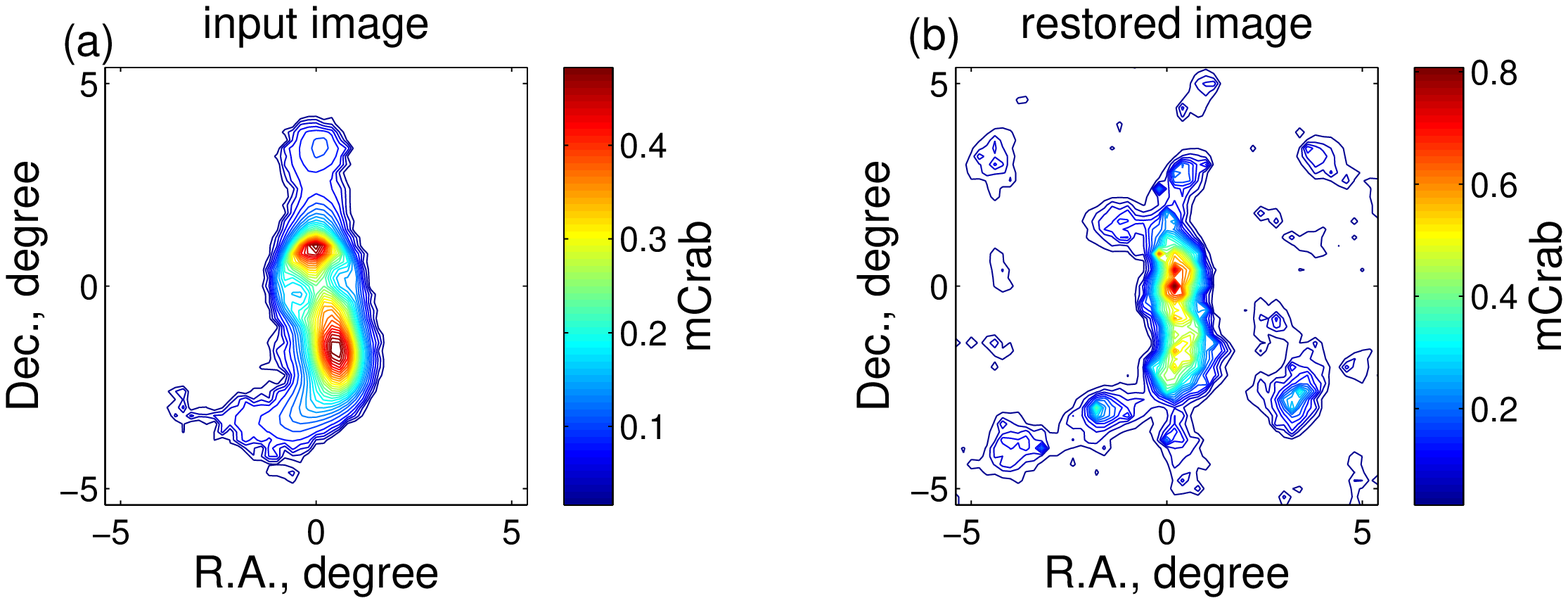}
		\figcaption{\label{fig9}   Case with the flux of input model decreased 5 times compared with Fig.~\ref{fig1}. (a) input image; (b) restored image by EMSME method. }
	\end{center}

	\begin{center}
		\includegraphics[width=9cm]{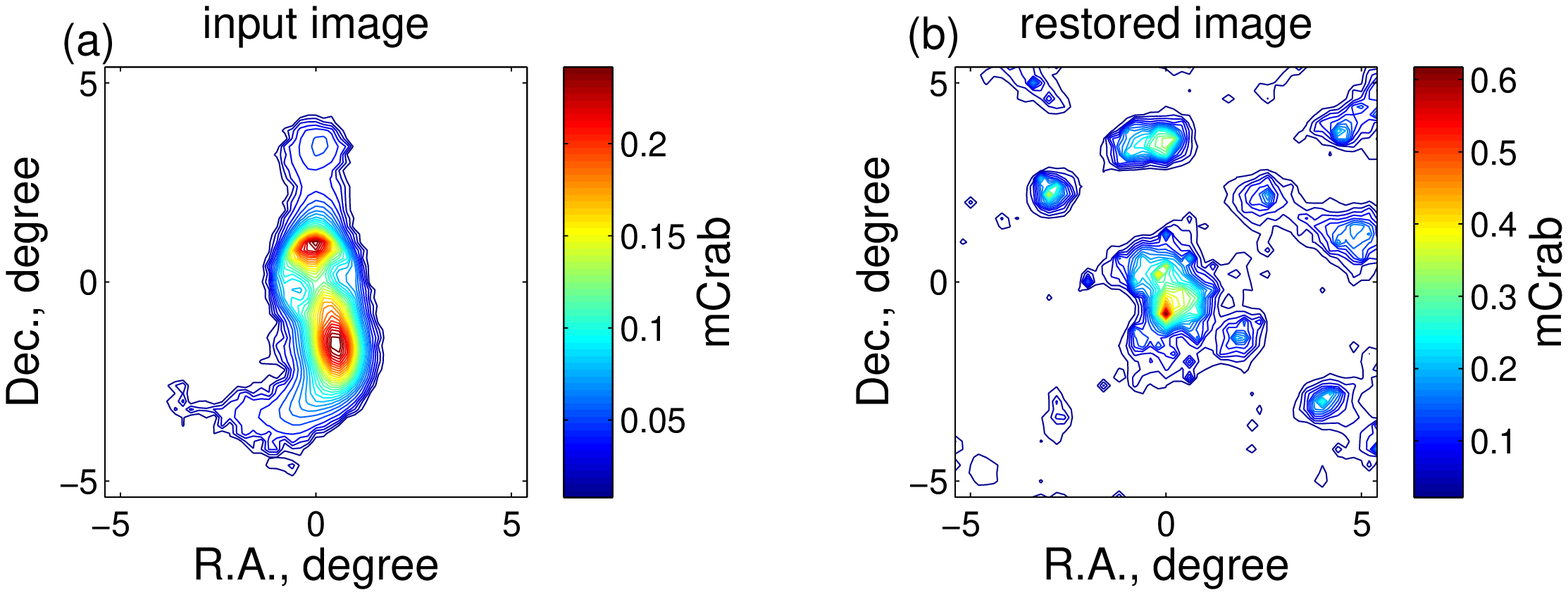}
		\figcaption{\label{fig10}  Case with the flux of input model decreased 10 times compared with Fig.~\ref{fig1}. (a) input image; (b) restored image by EMSME method.  }
	\end{center}

	\section{Conclusion and discussion}

	In this paper, we introduce a multiscale maximum entropy method with background constraint for the restoration of diffuse sources detected by HXMT imaging observation. This MSME method uses the wavelet transform to decompose an image into different frequency bands and includes noise modeling in the deconvolution. Regularized by the background constraint, it has a powerful denoising ability and is suitable for the restoration of diffuse source with different scale structures. It has been confirmed by simulation, which shows the restored image has a high correlation coefficient and Signal-to-Noise ratio.
	
	
	In order to overcome the mode-mixing problem in the MSME, an improved method is proposed, the ensemble multiscale maximum entropy (EMSME) method, which defines the final restored result as the mean of an ensemble of trials, each consisting of reconstructing the observed data plus a white noise of finite amplitude using the MSME method. Its principle is simple: the EMSME method uses the full advantage of the statistical characteristics of white noise, i.e. it perturbs the observed data and alleviates the coupling between signal and original noise, thus solving the mode-mixing to some extent, and cancels itself out in the space ensemble mean after serving its purpose. Simulation demonstrates that the EMSME method could further suppress noise and improve the correlation coefficient and SNR, and is thus better for image restoration.

	Indeed, this noise assisted data analysis technique could be applied to other restoration methods (such as the Richardson-Lucy algorithm and Direct Demodulation method) to help improve image restoration.
	
	Based on these two methods, HXMT could reach a limited detectivity of Cen A-like sources with a maximum differential flux of 0.5 mCrab in a one-time all-sky survey. Fainter sources could be detected by increasing the exposure time and the combined use of all the detectors with different fields of view onboard HXMT, which will be investigated in the future.

	The concept of background constraint is derived from the Direct Demodulation method and has been improved in this paper. We determine the background from different scale images through an iterative algorithm instead of the original image, hence it is more comprehensive. This improvement could be in turn used in the Direct Demodulation method, although this is beyond the scope of this paper.

\acknowledgments{The authors would like to thank Jing Jin for useful discussions about the observation modes of HXMT and are grateful to Fei Xie and Juan Zhang for offering the latest background simulation results of HXMT/HE. }

\vspace{15mm}

\end{multicols}

\vspace{-1mm}
\centerline{\rule{80mm}{0.1pt}}
\vspace{2mm}

\begin{multicols}{2}

\end{multicols}

\clearpage
\end{document}